
\input harvmac
\rightline{RU-93-51}
\vskip 1 in
\centerline{\bf Bound State Boundary S-matrix of the sine-Gordon Model}
\vskip .5 in
\centerline {Subir Ghoshal\foot{E-mail: GHOSHAL@ruhets.rutgers.edu}}
\centerline{Department of Physics and Astronomy}
\centerline{Rutgers University}
\centerline{P.O.Box 849, Piscataway, NJ 08855-0849}
\vskip 1 cm

\centerline{\bf Abstract}
 We study the boundary S-matrix for the reflection of bound states of
 the two-dimensional sine-Gordon integrable field theory in the presence of
 a boundary.
\lref\ghza{S.Ghoshal and A.B.Zamolodchikov,''Boundary S-matrix and
Boundary State in Two-dimensional Integrable Quantum Field
Theory'',Rutgers Preprint,RU-93-20,hep-th/9306002.}
\lref\zaza{A.B.Zamolodchikov,Al.B.Zamolodchikov.Ann.Phys.120(1979),253.}
\lref\tafa{L.Takhtadjian, L.Faddeev.Theor.Math.Phys.21,
(1974)p.160.}
\lref\kofa{V.Korepin, L.Faddeev.Theor.Math.Phys.25 (1975) p.147.}
\lref\clmn{S.Coleman.Phys.Rev.D11, (1975) p.2088.}
\lref\dveg{H.J.De Vega,A.Gonzalez Ruiz. Preprint LPTHE 92-45.}
\lref\frko{A.Fring, R.Koberle. ``Factorized Scattering in the Presence
of Reflecting Boundaries''. Preprint USP-IFQSC/TH/93-06, 1993.}
\lref\chrk{I.Cherednik. Theor.Math.Phys.,61,35 (1984) p.977.}
\newsec{\bf Introduction}
 In a recent paper \ghza, two-dimensional integrable field theory with a
 boundary has been studied. An essential ingredient of an integrable
 field theory in infinite space is the existence of an infinite number
 of mutually commuting integrals of motion. In the presence of the boundary,
 for arbitrary boundary conditions, these ``charges'' no longer remain
 conserved. However, sometimes, it is possible to modify these charges
 with special ``integrable'' boundary conditions so that the modified
 charges are indeed conserved. Then such a theory may be called a
 ``two-dimensional integrable field theory with a boundary''.
    \par An integrable ``bulk'' field theory enjoys the property that its
    multi-particle S-matrix amplitude factorizes into a product of an
    appropriate number of two-particle S-matrix amplitudes. The latter
    satisfy several constraints, namely, Yang-Baxter equation,
    unitarity and crossing symmetry \zaza. These constraints enable one to
    compute the exact S-matrix upto the so-called ``CDD''-factors. It has
    been known for quite some time how to generalise this factorizable
    structure of the S-matrix in the presence of a reflecting boundary
    \chrk. In addition to the ``bulk'' two-particle S-matrices one needs
    to introduce specific ``boundary reflection amplitudes'' for
    reflections of various particles in the theory off the boundary. The latter
have to
    satisfy appropriate generalisations of the constraints of the bulk
    theory - the boundary Yang-Baxter equation, the boundary unitarity
    condition and the boundary cross-unitarity condition. The last of
    these was introduced in \ghza. Thus one can in a way similar to the
    bulk case, pin down the factorizable boundary S-matrix, again, upto the
``CDD''-factors.
        \par In \ghza, the boundary S-matrices of the soliton scattering in
	the sine-Gordon model with a particular integrable boundary condition
	(which we call the ``boundary sine-Gordon model''), were obtained. In the
present work we
	compute the boundary S-matrix for reflections of the
	soliton-antisoliton bound states
	(the breathers) off the boundary. We employ the ideas of
	``boundary bootsrap'' as discussed in \refs{\ghza, \frko}.

\newsec{\bf Boundary S-matrix of the sine-Gordon Model}

In this section we review the study of the S-matrices of the sine-Gordon
model in the presence of the boundary \ghza .
 The bulk SG model is described by the action \zaza
,\eqn\ftac{\int_{-\infty}^{\infty}dy \int_{-\infty}^{\infty}dx\,
a(\phi,\partial_{\mu} \phi)} where, \eqn\sgac{a(\phi,\partial_{\mu}\phi
)={1\over
 2}(\partial_{\mu}\phi)^{2}-{m^{2}\over \beta^{2}}cos\beta\phi} where
 $\phi (x,y)$ is a scalar field and $\beta$ is a dimensionless coupling
 constant.The model is integrable both classically and quantum
 mechanically \refs{\tafa ,\kofa}. In the quantum theory, the discrete symmetry
$\phi
 \to \phi +{2\pi \over \beta}N,N\epsilon Z$ is spontaneously broken at
 $\beta^{2} < 8\pi$ \clmn; in this domain the theory is massive and its
 particle spectrum consists of a soliton-antisoliton pair $(A,\bar A)$ (with
equal masses) and a number (which depends on $\beta$) of
 neutral particles (``quantum breathers'')
 $B_{n}, n=1,2,...<\lambda$,where \eqn\lmda{\lambda={8\pi \over
 \beta^{2}}-1} The soliton (antisoliton) carries a positive (negative)
 unit of ``topological charge'' \eqn\chrg{q={\beta \over
 2\pi}\int_{-\infty}^{\infty} dx{\partial \over \partial x}\phi
 (x,y)={\beta \over 2\pi}[\phi(+\infty,y)-\phi (-\infty,y)] }The charge
 conjugation ${\bf C}:A \leftrightarrow \bar A$
 is related to
$\phi \leftrightarrow -\phi$ symmetry of \sgac. The particles
$B_n$ are neutral (they are interpreted as the soliton-antisoliton bound
states), $B_n$ with even (odd) $n$ being $\bf C$ -even ($\bf C$ -odd).
The masses of $B_n$ are

\eqn\bmas{m_n = 2M_{s}\sin({{n\pi}\over{2\lambda}}); \qquad n = 1, 2, ... <
\lambda,
}
where $M_s$ is the soliton mass.

Factorizable scattering of solitons is described by the commutation
relations $$A(\theta)A(\theta') = a(\theta-\theta')A(\theta')A(\theta), \qquad
\bar A(\theta)\bar A(\theta') =
a(\theta-\theta')\bar A(\theta')\bar A(\theta),$$ \eqn\cmre{ A(\theta)\bar
A(\theta') = b(\theta-\theta')\bar A(\theta')A(\theta) +
c(\theta - \theta')A(\theta')\bar A(\theta),}
where $A(\theta)$ and $\bar A(\theta)$ are soliton and antisoliton
creation operators and the two-particle scattering amplitudes $a, b, c$
are given by

$$a(\theta) = \sin(\lambda (\pi - u))\rho (u),$$
$$b(\theta) = \sin(\lambda u)\rho (u), $$
\eqn\ampl{c(\theta) = \sin(\lambda \pi)\rho (u),}
where $u = -i\theta$ and

$$\rho(u) = -{1\over\pi}\Gamma({\lambda})\Gamma(1-{{\lambda
u}/\pi})\Gamma(1-\lambda+{{\lambda
u}/\pi})\prod_{l=1}^{\infty}{{F_{l}(u)F_{l}(\pi - u)}\over
{F_{l}(0)F_{l}(\pi)}};$$
\eqn\rho{F_{l}(u) = {{\Gamma(2l\lambda -{\lambda u/\pi})\Gamma(1+2l\lambda
-{\lambda
u/\pi})}\over{\Gamma((2l+1)\lambda-{\lambda
u/\pi})\Gamma(1+(2l-1)\lambda-{\lambda u/\pi})}}.}
The amplitudes of $A B_n$ and $B_n B_m$ scatterings can be found in \zaza.
The amplitudes $b(\theta )$ and $c(\theta )$ have simple poles in the
``physical strip'', $Re \theta =0,\; 0 < Im \theta < \pi$, i.e. $0 < u <
\pi$ for \eqn\polb{u_{n}=\pi - {n\pi \over \lambda}, \qquad
n=1,2,... < \lambda}
These are interpreted as the neutral bound states $B_{n}$.
{}From the pole terms $$b(\theta) \simeq {f_{+-}^{n}f_{n}^{+-}
\over \theta -iu_{n}}$$ \eqn\verp{c(\theta) \simeq
{f_{+-}^{n}f_{n}^{-+} \over \theta - iu_{n}}} the vertices
$f_{ij}^{n}$ can be extracted and they are

$$f_{n}^{-+}=f_{n}^{+-}(-1)^{n}=f_{-+}^{n}$$ \eqn\vert{f_{+-}^{n}=f_{n}^{+-}}
The field theory in the presence of the boundary can be defined by
the following
action \foot{we disregard here, the possibility of the presence of additional
``boundary
degrees of freedom'' other than the boundary value of the ``bulk'' field
$\phi (x,y)$.}
\eqn\bdac{\int_{-\infty}^{\infty}dy \int_{-\infty}^{0}dx\; a(\phi
,\partial_{\mu}\phi )\; +\; \int_{-\infty}^{\infty}dy\,b(\phi_{B},{d\over
dy}\phi_{B})}
where $\phi_{B}=\phi(0,y)$. Based on an explicit computation of the first
non-trivial integral of motion, it was argued in \ghza, that the SG model is
integrable in the semi-infinite space with \eqn\inbc{b(\phi)=-M cos
({\beta \over 2}(\phi-\phi_{0}))} where $M$ and $\phi_{0}$ are free
parameters.The factorizable boundary scattering theory associated with
\inbc $\,$ was developed in \ghza .
The boundary S-matrix of the solitons and antisolitons can be
conveniently described by the following ``commutation relations'' between
soliton
and antisoliton creation operators $A(\theta)$ and ${\bar A}(\theta)$
and the formal ``boundary creation operator'' $B$ (see \ghza),

$$A(\theta)B = P_{+}(\theta)A(\theta)B + Q_{+}(\theta){\bar
A}(\theta)B;$$
\eqn\bamp{{\bar A}(\theta)B = P_{-}(\theta){\bar A}(\theta)B + Q_{-}(\theta)
A(\theta)B}
Here $P_{+}$, $Q_{+}$ ($P_{-}$, $Q_{-}$) are the amplitudes of
soliton (antisoliton) one-particle boundary scattering processes shown
in Fig.1. Exept for the case $M=\infty$, the boundary value $\phi(x=0,y)$ is
not fixed in the boundary
field theory \inbc $\,$ and hence the topological charge
\eqn\topc{q = {\beta \over{2\pi}}\int_{-\infty}^{0}dx
{\partial\over{\partial x}}\phi (x,y) }
is not conserved.That is why we allow processes described by $Q_{\pm}$.
The above amplitudes have to satisfy constraints resulting from the
boundary Yang-Baxter equation(BYB),the unitarity equation and the
cross-unitarity equation \ghza,described in  Fig.2, Fig.3 and Fig.4
respectively. The
BYB alone gives the solution \dveg ,\ghza,$$P_{+}(\theta) = \cos(\xi +\lambda
u)R(u);$$
$$P_{-}(\theta) = \cos(\xi -\lambda u)R(u);$$
$$Q_{+}(\theta) = {k_{+}\over 2}\sin(2\lambda u)R(u);$$
\eqn\samb{Q_{-}(\theta) = {k_{-}\over 2}\sin(2\lambda u)R(u),}
where again $u= -i\theta$; $\xi, k_{\pm}$ are free
parameters and $R(u)$ is an arbitrary function \foot{This is
the general solution for generic $\lambda$. For integer $\lambda$ there are
additional solutions.}.One can set $k_{+}=k_{-}=k$ by using a gauge
transformation
\eqn\guag{A(\theta) \to e^{i\alpha} A(\theta),  {\bar A}(\theta) \to
e^{-i\alpha}
{\bar A}(\theta) }
 The function
$R(u)$ can be determined using the ``boundary unitarity'' and the
``boundary cross-unitarity`` equations \ghza $$R(u)=R_{0}(u)R_{1}(u)$$where
$$R_{0}(u)=F_{0}(u)/F_{0}(-u);$$
$$F_{0}(u)={{\Gamma(1-{{2\lambda u}/\pi})}\over {\Gamma(\lambda
-{{2\lambda u}/\pi})}}\times$$
\eqn\rnot{\prod_{k=1}^{\infty}{{\Gamma(4\lambda k -
{{2\lambda u}/\pi})\Gamma(1 + 4\lambda k -{{2\lambda u}/\pi})
\Gamma(\lambda(4k+1))\Gamma(1+\lambda(4k-1))}
\over{\Gamma(\lambda(4k+1) -{{2\lambda u}/\pi})\Gamma(1 + \lambda
(4k-1) -{{2\lambda u}/\pi})\Gamma(1+4\lambda k)\Gamma(4\lambda k)}}} and

\eqn\rone{R_{1}(u) = {1\over {\cos\xi}}\sigma(\eta,u)\sigma(i\vartheta,u)}
where

$$\sigma(x,u) ={{\Pi(x, {\pi/2}-u)\Pi(-x, {\pi/2}-u)
\Pi(x, -{\pi/2}+u)\Pi(-x, -{\pi/2}+u)}\over{\Pi^{2}(x, {\pi/2})
\Pi^{2}(-x, {\pi/2})}};$$
$$\Pi(x, u) = \prod_{l=0}^{\infty}{{\Gamma(1/2 + (2l+1/2)\lambda +
{{x}/\pi} - {{\lambda u}/\pi})\Gamma(1/2 + (2l+3/2)\lambda +
{{x}/\pi})}\over{\Gamma(1/2 + (2l+3/2)\lambda +
{{x}/\pi} - {{\lambda u}/\pi})\Gamma(1/2 + (2l+1/2)\lambda +
{{x}/\pi})}} $$
solves

$$\sigma(x,u)\sigma(x,-u) = [\cos(x + \lambda u)\cos(x -
\lambda u)]^{-1};\quad \sigma(x,{\pi/2}-u)=\sigma(x,{\pi/2}+u), $$
and the parameters $\eta$ and $\vartheta$ are determined through the equations

\eqn\parm{\cos(\eta)\cosh(\vartheta) = - {1\over k}\cos\xi ; \quad \cos^2
(\eta) + \cosh^2 (\vartheta) = 1 + {1\over {k^2}}.}
\newsec{\bf Bound State Boundary S-matrix}
In this section we discuss the scattering of bound states $B_{n}$ of the
sine-Gordon model off the boundary.One starts with the commutation
relation (Fig.5) \eqn\bsam{B_{n}(\theta)B=R_{B}^{(n)}(\theta)B_{n}(-\theta)B}
where $B_{n}(\theta)$ creates the bound state $B_{n}$ with rapidity $\theta$.
The bound state boundary scattering amplitude $R_{B}^{(n)}(\theta)$
can be derived from the ``boundary bootsrap equation'' \refs{\ghza, \frko}
\eqn\bybe{f_{i_{1}i_{2}}^{n}(u_{n})R_{j_{1}}^{i_{1}}(u
+{u_{n}\over
2})S_{j_{2}f_{1}}^{i_{2}j_{1}}(2u)R_{f_{2}}^{j_{2}}(u-{u_{n}\over
2})=f_{f_{1}f_{2}}^{n}(u_{n})R_{B}^{(n)}(u)}where $R_{j}^{i}$ stand
for the amplitudes \samb $\;$ and $S_{ij}^{kl}(u)$ represent the two-particle
bulk S-matrices \ampl $\,$ with $u=-i\theta$. This equation is illustrated in
Fig.6. Furthermore
the solution must  satisfy the following:
\par 1.Boundary Unitarity Equation:
\eqn\bunt{R_{B}^{(n)}(u)R_{B}^{(n)}(-u)=1}This equation can be obtained as a
consistency condition by
applying \bsam $\,$twice. \par 2.Boundary Cross-unitarity Equation:
\eqn\bcun{R_{B}^{(n)}({\pi \over 2}-u)=R_{B}^{(n)}({\pi \over
2}+u)S^{(n,n)}(2u)} where $S^{(n,n)}(2u)$ is the scattering amplitude
for $B_{n}+B_{n} \to B_{n}+B_{n}$.These two conditions have been
discussed in detail in \ghza. Fig.3 and Fig.4 illustrate \bunt and \bcun
respectively if particles $a$ and $b$ are both taken to be $B_{n}$. The
equation \bybe $\,$ yields
$R_{B}^{(n)}(u)$  when the expressions \vert $\,$ for $f_{ij}^{n}$,
 \ampl $\,$ for $S_{ij}^{kl}(u)$ and \samb, \rnot - \parm $\,$ for
 $R_{j}^{i}$ are used. It can
be written as \eqn\bbsm{R_{B}^{(n)}(u)=R_{0}^{(n)}(u)R_{1}^{(n)}(u)}
where $$R_{0}^{(n)}(u)=(-1)^{n+1}{cos({u\over 2}+{n\pi\over
4\lambda})cos({u\over 2}-{\pi\over 4}-{n\pi\over 4\lambda})sin({u\over
2}+{\pi\over 4})\over cos({u\over 2}-{n\pi\over 4\lambda})cos({u\over
2}+{\pi\over 4}+{n\pi\over 4\lambda})sin({u\over 2}-{\pi\over
4})}$$ \eqn\brnt{\prod_{l=1}^{n-1}{sin(u+{l\pi\over 2\lambda})cos^{2}({u\over
2}-{\pi\over 4}-{l\pi\over 4\lambda})\over sin(u-{l\pi\over
2\lambda})cos^{2}({u\over 2}+{\pi\over 4}+{l\pi\over 4\lambda})}}
We find that $R_{1}^{(n)}(u)$, which contains the boundary parameters $\eta$
and
$\vartheta$ is different depending on whether $n$ is even or odd
\eqn\rbde{R_{1}^{(2n)}(u)=S^{(2n)}(\eta ,u)S^{(2n)}(i\vartheta ,u)}
where \eqn\sinr{S^{(2n)}(x,u)=\prod_{l=1}^{n}{sin (u)-cos({x\over
\lambda}-(l-{1\over 2}){\pi\over \lambda})\over sin(u)+cos({x\over
\lambda}-(l-{1\over 2}){\pi\over \lambda})} \; {sin(u)-cos({x\over
\lambda}+(l-{1\over 2}){\pi \over \lambda})\over sin(u)+cos({x\over
\lambda}+(l-{1\over 2}){\pi \over \lambda})}}$$n=1,2,.... < {\lambda
\over 2}$$ and \eqn\rbdo{R_{1}^{(2n-1)}(u)=S^{(2n-1)}(\eta
,u)S^{(2n-1)}(i\vartheta ,u)} with \eqn\snro{S^{(2n-1)}(x,u)={cos({x\over
\lambda})-sin(u)\over cos({x\over
\lambda})+sin(u)}\prod_{l=1}^{n-1}{sin(u)-cos({x\over \lambda}-{l\pi\over
\lambda})\over sin(u)+cos({x\over \lambda}-{l\pi\over
\lambda})} \; {sin(u)-cos({x\over \lambda}+{l\pi\over \lambda})\over
sin(u)+cos({x\over \lambda}+{l\pi\over \lambda})}} $$n=1,2,... <
{\lambda +1\over 2}$$ One can check that the solution so obtained
satisfies \bunt $\;$ and \bcun.

The factor $R_{0}^{(n)}(u)$ contains poles in the ``physical strip'' $0
< u < {\pi \over 2}$ located at $u={\pi \over 2}-{n\pi \over
2\lambda}$ for $\lambda > 1$. These poles can be explained as follows. In the
``direct
channel'' of the scattering $B_{n}+B_{n}\to B_{n}+B_{n}$ the
corresponding amplitude $S^{(n,n)}(u)$ shows a pole at $u={n\pi \over
\lambda}$ \zaza. This pole corresponds to the propagation of a real bound
state $B_{2n}$. As discussed in \ghza, the boundary state $|B > $
associated with the boundary condition \inbc $\,$ is expected to contain the
contributions of the zero-momentum particles $B_{2n}$. Therefore the
amplitude $R_{B}^{(n)}(u)$ must show a pole at $u={\pi\over
2}-{n\pi\over 2\lambda}$ as illustrated in Fig.7.
The pole at $u={\pi\over 2}$ is necessary to satisfy \bcun $\,$ at $u=0$
since $S^{(n,n)}(0)=-1$.
\par It is difficult to discuss the behavior of $R_{1}^{(n)}(u)$ for
    general values of the boundary parameters $\eta$ and $\vartheta$. As
    in \ghza, we will consider only two special cases:

a) Fixed Boundary Condition:
  In this case \rone $\,$ becomes \ghza,
\eqn\rfix{R_{1}(u)={1\over cos\xi}\sigma(\xi,u)} and consequently we get
\eqn\rbfx{R_{1}^{(n)}(u)=S^{(n)}(\xi,u)}There exist poles in the
``physical strip'' $0 < u < {\pi\over 2}$ in $R_{1}^{(n)}(u)$ and these
represent some ``boundary bound states''.

b) Free Boundary Condition:
  In this case $\eta = {\pi\over 2}(\lambda +1)$ and $\vartheta
  =0$  \ghza. There exists a pole in $R_{1}^{(n)}(u)$ for both even and odd
  values of $n$ at $u={n\pi\over 2\lambda}$. This pole corresponds to a
  ``boundary bound state'' propagating along the boundary as illustrated
  in Fig.8. As discussed in \ghza, the energy of this ``boundary bound state''
should be
  $e_{n}=e_{0}+m_{n}cos({n\pi\over 2\lambda})$, where $e_{0}$ is the
  ground state energy. Using \bmas, we get
  $e_{n}=e_{0}+M_{s}sin({n\pi\over \lambda})$. The existence of this set
  of bound states with energies {$e_{n}$} can be seen also from the
  soliton boundary S-matrices \samb. For the ``free'' boundary
  conditions, $P_{+}(\theta)=P_{-}(\theta)=P_{free}(\theta)$ and
  $Q_{+}(\theta)=Q_{-}(\theta)=Q_{free}(\theta)$ both show poles at
  $\theta =i{\pi\over 2}-i{n\pi\over \lambda}$. Energies of the
  corresponding ``boundary bound states'', when computed, agree with
  {$e_{n}$} found above.
\vskip .5 cm
\centerline{Acknowledgement}
 I would like to thank A.B.Zamolodchikov for advice, inspiration and
 numerous illuminating discussions.

\listrefs
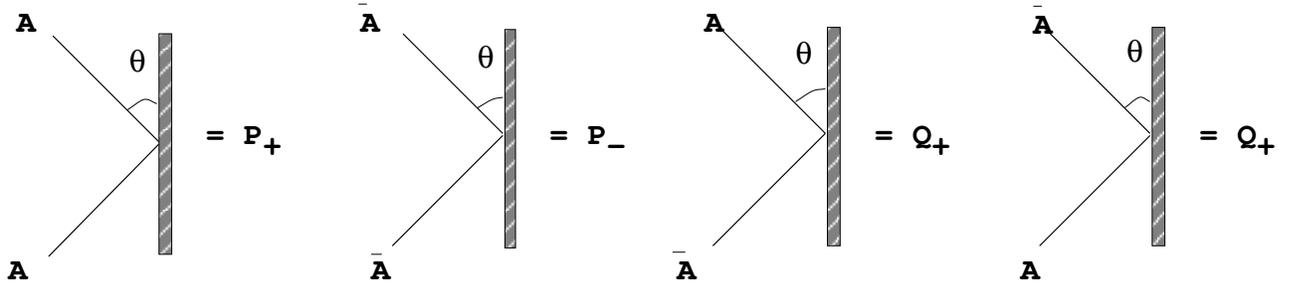
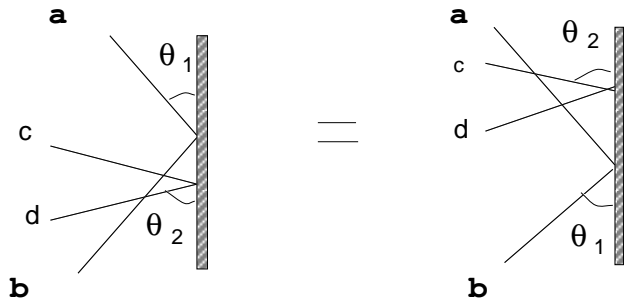
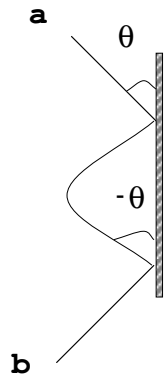
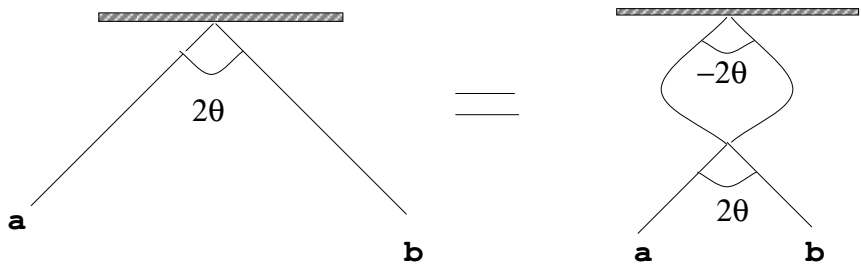
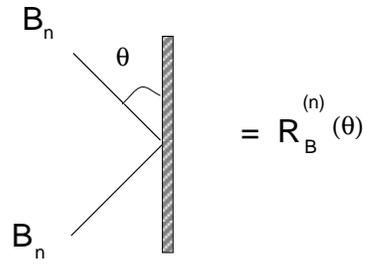
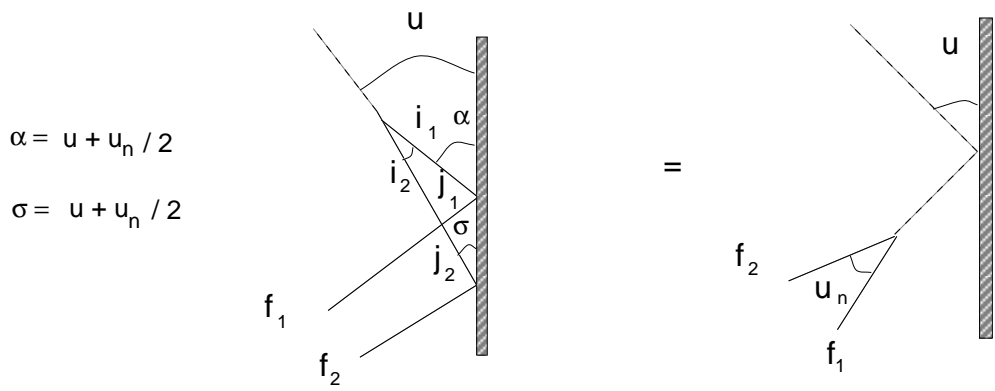
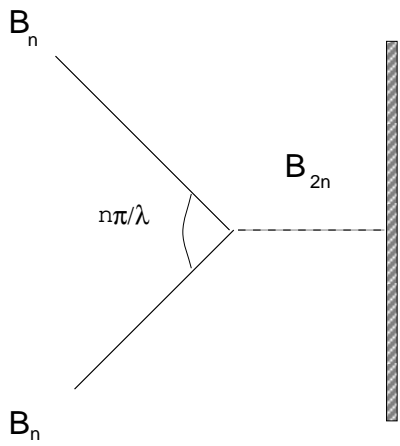
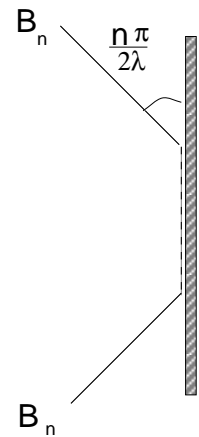
\end